\documentclass[manuscript]{emulateapj}

\usepackage{amssymb}
\usepackage[export]{adjustbox}
\usepackage{amsmath}
\usepackage{booktabs}
\usepackage{hyperref}
\usepackage{multirow}
\usepackage{times}
\PassOptionsToPackage{hyphens}{url}\usepackage{hyperref}
\usepackage[hyphenbreaks]{breakurl}

\newcommand{\xmm}{{\it XMM-Newton}}

\newcommand{\arcs}{\hbox{$^{\prime\prime}$}}

\newcommand{\ls}
{\mathrel{\hbox{\rlap{\hbox{\lower4pt\hbox{$\sim$}}}\hbox{$<$}}}}
\newcommand{\gs}
{\mathrel{\hbox{\rlap{\hbox{\lower4pt\hbox{$\sim$}}}\hbox{$>$}}}}

\received{}
\begin{document}
\title{Resolving the soft X-ray ultra fast outflow in PDS 456}
\shorttitle{The soft X-ray outflow in PDS 456.}
\shortauthors{Reeves et al.}
\author{J. N. Reeves\altaffilmark{1,2}, V. Braito\altaffilmark{1,2}, G. Chartas\altaffilmark{3}, F. Hamann\altaffilmark{4}, S. Laha\altaffilmark{1,5}, E. Nardini\altaffilmark{6,7}} 
\altaffiltext{1}{Center for Space Science and Technology, University of Maryland Baltimore County, 1000 Hilltop Circle, Baltimore, MD 21250, USA; email jreeves@umbc.edu}
\altaffiltext{2}{INAF, Osservatorio Astronomico di Brera, Via Bianchi 46 I-23807 Merate (LC), Italy}
\altaffiltext{3}{Department of Physics and Astronomy College of Charleston, Charleston, SC, 29424, USA}
\altaffiltext{4}{Department of Physics \& Astronomy, University of California, Riverside, CA 92507, USA}
\altaffiltext{5}{Astroparticle Physics Laboratory, NASA/Goddard Spaceflight Center, Mail Code 661, Greenbelt, MD 20771, USA}
\altaffiltext{6}{INAF, Osservatorio Astrofisico di Arcetri, Largo Enrico Fermi 5, I-50125 Firenze, Italy}
\altaffiltext{7}{Dipartimento di Fisica e Astronomia, Universit\`a di Firenze, via G. Sansone 1, I-50019 Sesto Fiorentino, Firenze, Italy}

\begin{abstract}

Past X-ray observations of the nearby luminous quasar PDS 456 (at $z=0.184$) have revealed a wide angle accretion disk wind \citep{Nardini15}, with an outflow velocity of $\sim-0.25c$, as observed through observations of its blue-shifted iron K-shell absorption line profile. 
Here we present three new {\it XMM-Newton} observations of PDS\,456; one in September 2018 where the quasar was bright and featureless, 
and two in September 2019, 22 days apart, occurring when the quasar was five times fainter and where strong blue-shifted lines from the wind were present. 
During the second September 2019 observation, three broad ($\sigma=3000$\,km\,s$^{-1}$) absorption lines were resolved in the high resolution RGS spectrum, 
which are identified with blue-shifted O\,\textsc{viii} Ly$\alpha$, Ne\,\textsc{ix} He$\alpha$ and Ne\,\textsc{x} Ly$\alpha$. 
The outflow velocity of this soft X-ray absorber was found to be $v/c=-0.258\pm0.003$, fully consistent with iron K absorber with $v/c=-0.261\pm0.007$. 
The ionization parameter and column density of the soft X-ray component ($\log\xi=3.4$, $N_{\rm H}=2\times10^{21}$\,cm$^{-2}$) outflow was lower by about two orders of magnitude, when compared to the high ionization wind at iron K ($\log\xi=5$, $N_{\rm H}=7\times10^{23}$\,cm$^{-2}$). 
Substantial variability was seen in the soft X-ray absorber between the 2019 observations, declining from 
$N_{\rm H}=10^{23}$\,cm$^{-2}$ to $N_{\rm H}=10^{21}$\,cm$^{-2}$ over 20 days, while the iron K component was remarkably stable. We conclude that the soft X-ray wind may originate from an inhomogeneous wind streamline passing across the line of sight and which due to its lower ionization, is located further from the black hole, 
on parsec scales, than the innermost disk wind.
\end{abstract}









\section{Introduction}

The discovery of blueshifted absorption features from K-shell transitions of iron has provided strong evidence for the presence of high-velocity outflows in AGN. The measured blueshifts imply outflow velocities of the order of $\sim-0.1c$ or higher (e.g. APM\,08279+5255; \citealt{Chartas02}, PG\,1211+143; \citealt{Pounds03}, PDS\,456; \citealt{Reeves03}).  The high velocities and high ionization of the material (from He and H-like Fe) suggest that the high-velocity outflows originate from much closer to the black hole than the slower, less-ionized warm absorbers (e.g. \citealt{Kaastra00, Kaspi02, Crenshaw03, McKernan07}).
Systematic studies of archival {\it XMM-Newton} \citep{Tombesi10, Tombesi11} and {\it Suzaku} \citep{Gofford13} data, as well as from X-ray variability studies \citep{Igo20}, 
have shown that high-velocity iron K winds may be a common feature of nearby AGN.  The derived outflow rates are calculated to be high (up to $\sim$few $M_{\odot}$\,yr$^{-1}$) and therefore comparable with the measured accretion rates of AGN, while the outflows are estimated to carry kinetic power as much as a few per cent of the bolometric luminosity, $L_{\rm bol}$ \citep{Tombesi12,Gofford15}.  
As such, they may play an important role in linking black-hole growth and the properties of the host galaxy \citep{King03, King10} and offer a possible interpretation of the $M-\sigma$ relation for galaxies \citep{FerrareseMerritt00, Gebhardt00}.

While the evidence for ultra fast outflows in the Fe\,K band is now well established, there has been a relative scarcity of detections to date in the soft X-ray band.  Detecting lower-ionization, soft-band counterparts of high-velocity outflows is important in order to probe all phases of the outflowing gas and its structure across a wide range of ionization and column density.
Indeed, the original evidence for the fast wind seen in PG\,1211+143 \citep{Pounds03} came from both its iron K profile and from blue-shifted soft X-ray lines in the 
{\it XMM-Newton} RGS spectrum and subsequently confirmed in analysis of later datasets \citep{Pounds16, RLP18}. 

During an initial {\it XMM-Newton} observation of the nearby QSO PDS\,456 in 2001, \citet{Reeves03} first noted the presence of blue-shifted absorption in the soft X-ray band. A broad absorption trough was resolved in the RGS spectrum, attributed to a blend of L-shell transitions from highly-ionized Fe near to $\sim$1\,keV, with an estimated  outflow velocity of $\sim -50\,000$\,km\,s$^{-1}$, in addition to the now well-established high-velocity outflow detected in the Fe\,K band 
\citep{Reeves09, Nardini15, Matzeu17a}, where the outflow velocity has been confirmed to vary over a narrow range between $0.25-0.30c$.  
The presence of high-velocity, soft-band counterparts of the outflow in PDS 456 have since been confirmed in a multi-epoch analysis of all of the archival {\it XMM-Newton} observations of PDS\,456 \citep{Reeves16}, although the soft X-ray velocities did not always match those measured in the iron K-shell band. 

Concerning other AGN, a soft-band absorber with a velocity of $\sim -0.25c$ was also observed in IRAS\,13224$-$3809 through high-resolution RGS data \citep{Pinto18}, where the  corresponding highly-ionized Fe\,K counterpart has a similar velocity as per the soft X-ray lines \citep{Parker17}.  
The narrow line Seyfert 1 galaxy, 1H\,0707$-$495, which has a very similar nature to IRAS\,13224$-$3809, also appears to show a variety of blueshifted absorption line features associated with a fast wind with {\it XMM-Newton}, at a velocity of $\sim0.13c$ \citep{Kosec18}. Curiously this AGN also displays soft X-ray emission lines which are 
blue-shifted by up to 8000\,km\,s$^{-1}$, which \citet{Kosec18} interpret in the framework of a slowing, cooling outflow on larger scales.  
A fast soft X-ray outflow was observed in the {\it XMM-Newton} RGS spectrum of the narrow-line Seyfert 1 galaxy IRAS\,17020+4544 \citep{Longinotti15}, covering a wide range of ionization and column density with outflow velocities in the range $-23\,000$ to $-33\,000$\,km\,s$^{-1}$, although an iron K-shell counterpart to this wind has not yet been reported. 
Further claims of soft X-ray fast wind detections noted in the literature include Ark\,564 \citep{Gupta13} and Mrk\,590 \citep{Gupta15}, in these cases performed with the gratings on-board {\it Chandra}, 
although with no known iron K-shell component.
Thus while the evidence for soft X-ray components of ultra fast outflows is growing, the exact relationship between the iron K and soft X-ray fast winds is still poorly understood

Here, we present three new observations of PDS\,456 from 2018--2019, taken with {\it XMM-Newton}, of which two were also simultaneous with {\it NuSTAR}, which are 
described further below. 
As discussed above, the radio-quiet quasar PDS\,456 (at $z=0.184$, \citealt{Torres97}) was one of the first prototype examples of an ultra fast outflow.
Indeed, since its initial detection in 2001 with \xmm\ \citep{Reeves03}, the presence of the ultra fast outflow in PDS\,456 has now been established through over a decade's worth of X-ray observations \citep{Reeves09,Behar10,Reeves14,Gofford14,Nardini15,Hagino15,Matzeu16,Matzeu17a,Matzeu17b,Parker18,Reeves18a,Reeves18b,BM19}. 
Furthermore \citet{Hamann18} recently claimed a fast UV counterpart to the X-ray wind 
on the basis of a broad C\,\textsc{iv} trough predicted from photoionization modeling.  

In this paper we reveal the structure of the wind through the detection of resolved soft X-ray absorption lines in the {\it XMM-Newton} RGS, which as we will show, exactly match the outflow velocity of the wind as measured in the iron K band. Furthermore, from the wind variability, we also probe the location and properties of the soft X-ray wind component. 

\section{Observations and Data Reduction}

PDS 456 was observed a further three times with  \xmm\ over 2018--2019, as part of a new campaign to study the long-term variability of its X-ray ultra fast outflow and 
to study its possible connection to the UV wind. A summary of the overall campaign is listed in Table\,1. Note that the observation in September 2018 was performed simultaneously with {\it NuSTAR} and {\it HST}, the 2nd observation in early September 2019 
(hereafter 2019a) was coordinated with {\it HST}, while the 3rd observation in late September 2019 (hereafter 2019b) was simultaneous with {\it NuSTAR}. 
All {\it XMM-Newton} observations were performed in Large Window mode in the pn and MOS to mitigate any photon pile-up.

The X-ray observations were processed using the \textsc{nustardas} v1.8.0, 
{\it XMM-Newton} \textsc{sas} v18.0 and \textsc{heasoft} v6.25 software.
{\it NuSTAR} source spectra were extracted using a 50\arcs\ circular region centered on the source and background from a 76\arcs\ circular region clear from stray light. 
{\it XMM-Newton} EPIC-pn spectra were extracted from single and double events, using a 30\arcs\ source region and $2\times34\arcs$\ background regions on the same chip. 
The spectra and responses from the individual MOS\,1 and MOS\,2 CCDs were combined into a single spectrum after they were first checked for consistency, as were the 
individual spectra from the FPMA and FPMB detectors on-board {\it NuSTAR}.  
The {\it NuSTAR} and {\it XMM-Newton} pn and MOS spectra are binned to at least 50 counts per bin. 
Spectra from the {\it XMM-Newton} Reflection Grating Spectrometer (RGS, \citealt{denHerder01}) were extracted using the \textsc{rgsproc} pipeline and were combined into a single spectrum for each of the observations, after first checking that the individual 
RGS 1 and RGS 2 spectra were consistent with each other within the errors. 

\begin{deluxetable}{lccc}
\tablecaption{Log of 2018--2019 PDS 456 {\it XMM-Newton} Observations.}
\tablewidth{0pt}
\tablehead{
& \colhead{2018} & \colhead{2019a} & \colhead{2019b}}
\startdata
OBSID & 0830390101 & 0830390201 & 0830390401\\
Start date & 2018/09/20 & 2019/09/02 & 2019/09/24\\
Start time (UT) & 13:18:00 & 15:24:15 & 13:53:32 \\
Duration$^{a}$ & 86.0 & 83.0 & 94.3 \\  
Net rate$^{b}$ & $3.958\pm0.008$ & $0.821\pm0.004$ & $1.254\pm0.004$\\
Flux$^{c}$ & 9.43 & 1.87 & 2.24 \\
Coverage$^{d}$ & {\it HST} + {\it NuSTAR} & {\it HST} & {\it NuSTAR} 
\enddata
\tablenotetext{a}{Total duration of exposure in ksec with {\it XMM-Newton} EPIC-pn.}
\tablenotetext{b}{Net count rates (cts\,s$^{-1}$), over the 0.4--10\,keV band for {\it XMM-Newton} EPIC-pn.}
\tablenotetext{c}{Flux in 2-10\,keV band compared to an absorbed power-law, in units $\times10^{-12}$\,ergs\,cm$^{-2}$\,s$^{-1}$.}
\tablenotetext{d}{Coordinated with either {\it HST/COS} or {\it NuSTAR} (or both). All observations have {\it Swift} coverage.}
\label{tab:obslog}
\end{deluxetable}

\subsection{Overview of the New Campaign}

In order to provide an overview of the whole campaign, Figure\,1 (left panel) shows the broad band {\it XMM-Newton} EPIC-pn and {\it NuSTAR} X-ray spectra obtained for the three observations. The observations captured significant variability of the QSO. The 2018 observation caught PDS\,456 in a very bright state (where $F_{2-10}=9.4\times10^{-12}$\,ergs\,cm$^{-2}$\,s$^{-1}$), following a major X-ray flare which occurred during the {\it Swift} monitoring. The X-ray spectrum of this epoch is continuum dominated, while the soft X-ray spectrum (as seen by the RGS) is featureless. Details of the 2018 observation, the possible presence of a high energy cut-off in the {\it NuSTAR} spectrum, the X-ray flare as seen by concurrent {\it Swift} monitoring and the inferred X-ray coronal properties of PDS\,456 will be discussed in a subsequent paper. 

In comparison, the two 2019 observations are a factor of $\times 5$ fainter over the 2--10\,keV band compared to the bright 2018 observation. Both 2019 observations show similar pronounced iron K absorption
over the 7--10\,keV band. Of these, the 2019a spectrum is more obscured at soft X-rays and is about 50\% fainter than the 2019b spectrum below 2\,keV. Indeed Figure\,1 (right panel) 
shows the {\it Swift} X-ray Telescope (XRT) lightcurve over the 0.3--10 keV band obtained during the 2019 campaign, where the 2019a {\it XMM-Newton} 
observation occurred near a period of minimum flux. The 2019b \xmm\ observation is at a typical flux level for the 2019 monitoring campaign, although this is 
still at a much lower flux than in 2018.

\begin{figure*}
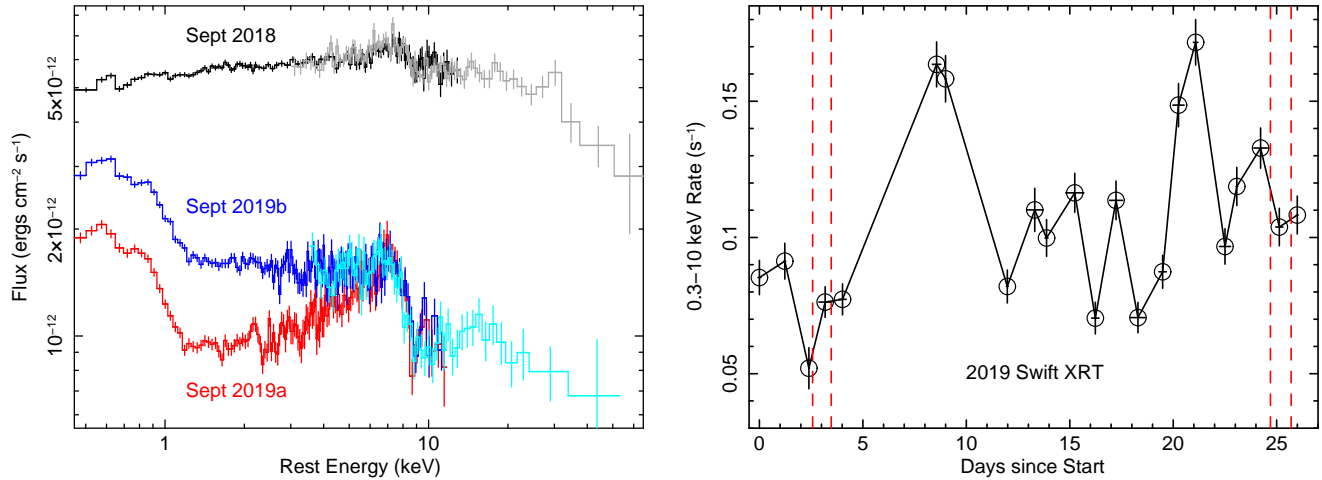

\begin{center}
\rotatebox{270}{\includegraphics[height=8.9cm]{f1a.eps}}
\rotatebox{270}{\includegraphics[height=8.9cm]{f1b.eps}}
\end{center}
\caption{Left panel. Broad-band {\it XMM-Newton} and {\it NuSTAR} X-ray spectra of the three observations of PDS\,456, observed from September 2018 to September 2019 (see Table 1 for details). 
The September 2018 spectrum is plotted in black (greyscale for {\it NuSTAR}), the first of the September 2019 spectra in red (2019a) and the second September observation (2019b) 
in blue (cyan for {\it NuSTAR}). Note the spectra are plotted against a simple $\Gamma=2$ power-law to create $\nu F_{\nu}$ spectra and are corrected for the Galactic absorption column. The campaign captured strong variability from PDS\,456. 
The 2018 observation was observed in a high flux state, which followed a major flare as seen during monitoring with {\it Swift}. Its X-ray spectrum is mainly featureless, except for a roll-over present above 10 keV 
seen in {\it NuSTAR} due to a high energy cut-off. In contrast the 2019 spectra are at least a factor of $\times 5$ lower in flux and show pronounced absorption at iron K in the 7-10 keV band, while the 2019a spectrum also shows strong soft X-ray absorption. Right panel. 2019 {\it Swift} XRT lightcurve, performed from 2019/08/31 to 2019/09/26. 
The vertical dashed lines show the intervals of the two 2019 {\it XMM-Newton} pointings, where the 2019a observation occurred near a minimum in the lightcurve.}
\label{fig:compare}
\end{figure*}

In this paper, we concentrate our initial analysis on the 2019b observation, while we also compare the results to the spectra obtained in the 2018 and 
2019a epochs to test the wind variability. 
As we will subsequently show, the 2019b observation captured PDS 456 in a wind dominated state where we resolve 
multiple absorption line profiles from the outflow in both the soft X-ray band with \xmm\ RGS and at iron K in the broad-band spectrum. 
In contrast the 2019a observation caught the source in a more obscured state, where its very low 
soft X-ray flux prohibits a detailed analysis of the RGS spectrum. The simultaneous HST/COS spectroscopy of PDS\,456 (PI F. Hamann), 
coincident with the 2018 and 2019a observations, 
where the motivation was to further monitor the outflowing UV absorption towards PDS\,456, will be presented in later work. 
Note that outflow velocities are given with respect to the rest-frame of the host galaxy of PDS\,456 at $z=0.184$, 
after correcting for relativistic Doppler shifts. Errors are quoted at 90\% confidence for one interesting parameter (or $\Delta\chi^2=2.7$).

\section{Soft X-ray Spectral Analysis}

\subsection{The XMM-Newton RGS Spectrum}\label{RGS}

We first analyze the soft X-ray spectrum of PDS 456 from the 2019b observation, obtained with the {\it XMM-Newton} RGS. The total exposure of the 
RGS observation is 92.4\,ks, while the net count rate for the combined RGS\,1+2 spectrum is $0.070\pm0.001$\,cts\,s$^{-1}$, yielding $\sim 6500$ net counts. 
The RGS spectrum was grouped into bins of width $\Delta\lambda=0.08$\,\AA, so that each spectral bin corresponds to the approximate 
FWHM spectral resolution of the RGS. At an observed wavelength of 16\,\AA, this equates to a resolution of $\lambda/\Delta\lambda=200$ (or 
$\sim1500$\,km\,s$^{-1}$ FWHM).
A further binning was applied to achieve a minimum signal to noise of 3 per bin, this 
ensures that the extreme low and high wavelength ends of the spectrum are not too noisy, in order for the continuum to be well defined.
The RGS spectra were analyzed over the $6-25$\,\AA\ observed wavelength range (or $5-21$\,\AA\ QSO 
rest frame); at longer wavelengths the spectrum is suppressed by Galactic absorption.  
The latter was included in the spectral fitting, using the \textsc{tbabs} model of \citet{Wilms00}, where for PDS 456 the column density is expected to be $N_{\rm H} = 2.4 \times 10^{21}$\,cm$^{-2}$ based on 21 cm measurements \citep{Kalberla05}. Solar abundances of \citet{GRSA98} and a
conversion between energy and wavelength of $1\,{\rm keV} = 1.602\times10^{-9}\,{\rm ergs} = 12.3984$\,\AA\ were adopted. 

\begin{figure*}
\begin{center}
\rotatebox{-90}{\includegraphics[height=15cm]{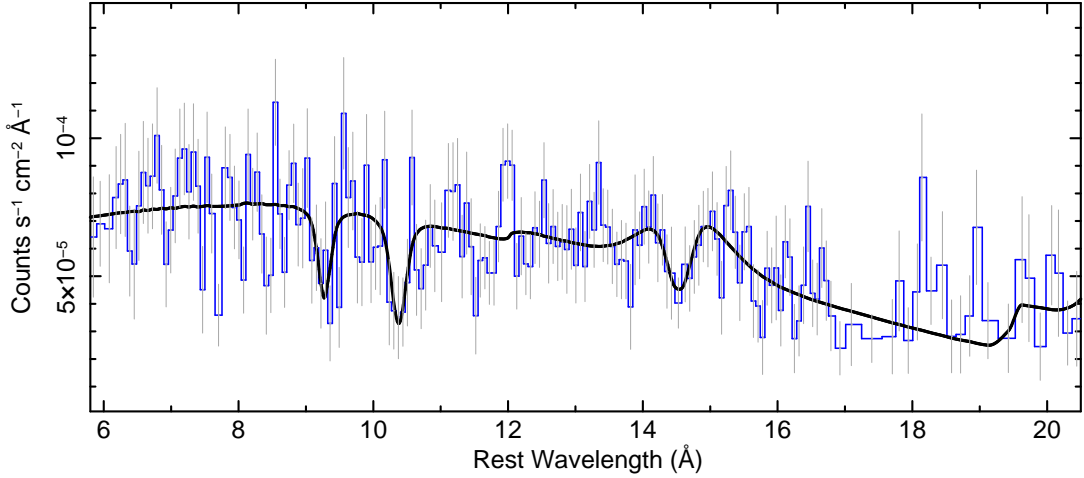}}
\end{center}
\caption{Count rate {\it XMM-Newton} RGS spectrum of PDS\,456 from the 2019b observation, plotted in the QSO rest frame. $1\sigma$ error bars are shown in grey. The continuum model (black line) consists of a powerlaw (photon index $\Gamma=2.71\pm0.09$) absorbed by the Galactic column. 
The spectrum shows three significant absorption lines modeled by Gaussian profiles, centered at rest wavelengths 
of $\lambda=14.56\pm0.05$\,\AA,  $\lambda=10.37\pm0.04$\,\AA\ and $\lambda=9.30\pm0.06$\,\AA\ (see Table~2). 
The lines could be associated to the predicted strong resonance ($1s\rightarrow2p$) transitions of O\,\textsc{viii} Ly$\alpha$ (18.97\,\AA), Ne\,\textsc{ix} He$\alpha$ (13.45\,\AA) and Ne\,\textsc{x} Ly$\alpha$ (12.13\,\AA), 
with a common velocity shift of $v/c=-0.257\pm0.003$ from a fast wind. Note that the decline in flux long-wards of 16\,\AA\ is due to Galactic absorption.}
\label{fig:rgs}
\end{figure*}

A model consisting of a power-law absorbed by the Galactic column returned a poor fit, with a reduced chi-squared of $\chi^{2}_{\nu}=334.4/254$ and a photon index of $\Gamma=2.71\pm0.09$. 
The RGS spectrum is plotted in Figure 2 and three possible absorption lines appear to be present against the power-law continuum. 
These were fitted with Gaussian profiles with rest-frame centroid wavelengths (energies) of $\lambda=14.56\pm0.05$\,\AA\ (or $E=851.8\pm3.5$\,eV), $\lambda=10.37\pm0.04$\,\AA\ ($E=1196\pm5$\,eV)
and $\lambda=9.30\pm0.06$\,\AA\ ($E=1333\pm9$\,eV). The line parameters are summarized  in Table~2. 
The lines appear to be significant when added to the model; the overall fit-statistic improves by $\Delta\chi^2=-28.9$, $\Delta\chi^2=-19.0$ 
and $\Delta\chi^2=-9.3$ respectively upon the addition of each line with two extra degrees of freedom. 

\begin{deluxetable*}{lccccc}
\tablecaption{Blueshifted absorption line parameters to the 2019b PDS 456 observation.}
\tablewidth{0pt}
\tablehead{\colhead{ID} 
& \colhead{O\,\textsc{viii} Ly$\alpha$} & \colhead{Ne\,\textsc{ix} He$\alpha$} & \colhead{Ne\,\textsc{x} Ly$\alpha$} & \colhead{Fe\,\textsc{xxvi} Ly$\alpha$} & \colhead{Fe\,\textsc{xxvi} Ly$\alpha$}}
\startdata
Instrument & RGS & RGS & RGS & EPIC-pn & EPIC-pn \\
$\lambda_{\rm rest}$ or $E_{\rm rest}$$^{a}$ & $14.56\pm0.05$\,\AA & $10.37\pm0.04$\,\AA & $9.30\pm0.06$\,\AA & $8.98\pm0.13$\,keV & $10.2\pm0.2$\,keV\\
$EW ({\rm eV})$$^{b}$ & $-11.5^{+2.9}_{-3.2}$ & $-16.2^{+4.6}_{-5.3}$ & $-14.0^{+5.0}_{-6.0}$ & $-255^{+58}_{-75}$ & $-246\pm70$\\
$\sigma ({\rm eV})^{c}$ & $8.6^{+2.7}_{-2.1}$ & $12.1^{t}$ & $13.5^{t}$ & $310^{+140}_{-90}$ & $310^f$\\
$\sigma_{\rm v}$ (km\,s$^{-1}$)$^{d}$ & $3030^{+950}_{-740}$ & $3030^{t}$ & $3030^{t}$ & $10300^{+4700}_{-3000}$\\
$\lambda_{\rm lab}$ or $E_{\rm lab}$$^{e}$ & 18.97\,\AA & 13.45\,\AA & 12.13\,\AA & 6.97\,keV & 6.97\,keV \\
$v/c^{g}$ & $-0.259\pm0.003$ & $-0.255\pm0.005$ & $-0.260\pm0.009$ & $-0.25\pm0.02$ & $-0.36\pm0.02$ \\
$\Delta\chi^{2}/\Delta\nu$$^{h}$ & $-28.9/3$ & $-19.0/2$ & $-9.4/2$ & $-49.6/3$ & $-30.1/2$
\enddata
\tablenotetext{a}{Rest wavelength (RGS) in \AA\ or energy (EPIC-pn) in keV of absorption line in the observed spectrum.}
\tablenotetext{b}{Equivalent width of absorption line in units of eV.}
\tablenotetext{c}{$1\sigma$ Gaussian width of line in units of eV.}
\tablenotetext{d}{$1\sigma$ velocity width of line in units of eV.}
\tablenotetext{e}{Expected lab-frame wavelength (or energy) for above line ID, with no velocity shift.}
\tablenotetext{f}{Indicates parameter is fixed.}
\tablenotetext{g}{Blueshift of absorption line in units of c for above line ID.}
\tablenotetext{g}{Improvement in fit statistic ($\Delta\chi^2$), for the addition of $\Delta\nu$ degrees of freedom to the model.}
\tablenotetext{t}{Indicates line velocity width is tied with respect to that of the O\,\textsc{viii} Ly$\alpha$ line (in velocity space).}
\label{tab:lines}
\end{deluxetable*}

We first test the scenario whereby the lines arise from a soft X-ray component of the known fast wind in PDS\,456. 
Multiple X-ray observations of PDS\,456 spanning more than a decade have shown the frequent presence of a fast outflow at iron K, with a typical velocity of $0.25-0.3c$ 
\citep{Reeves09, Nardini15, Matzeu17a}, while the possible presence of blue-shifted soft X-ray features have also been noted in prior RGS spectra \citep{Reeves16}. 
In this scenario, the 14.56\,\AA\ line could be identified with blueshifted O\,\textsc{viii} Ly$\alpha$ (at 18.97\AA) and the derived velocity shift is 
then $v/c=-0.259\pm0.003$. Similarly, identifying the shorter wavelength lines with the strong He and H-like ($1s\rightarrow2p$) lines of Ne\,\textsc{ix} He$\alpha$ and Ne\,\textsc{x} Ly$\alpha$ (at lab-frame wavelengths of 13.45\,\AA\ and 12.13\,\AA) gives consistent velocity shifts, with $v/c=-0.255\pm0.005$ and $-0.260\pm0.009$ respectively. If a common velocity shift is assumed for all three lines, then the derived outflow velocity is $v/c=-0.257\pm0.003$. In the following sub-section these identifications will be placed on a firmer footing,  
based on the photoionization analysis. 

Regardless of the identifications, the absorption line profiles are resolved in the RGS spectrum. A zoom-in of the three line profiles are shown in Figure~3, where each spectral bin corresponds to the FWHM resolution of the RGS gratings. Assuming a common velocity broadening across all three lines gives a velocity width of $\sigma_{\rm v}=3030^{+950}_{-740}$\,km\,s$^{-1}$ (or a FWHM of $\sim7000$\,km\,s$^{-1}$). The velocity widths of the individual lines are also consistent within errors if they are allowed to vary independently of each other.  For the 14.56\,\AA\ line, the $1\sigma$ width is then $\Delta\lambda=0.16^{+0.07}_{-0.05}$\,\AA, which 
corresponds to a velocity width of $\sigma_{\rm v}=3400^{+1300}_{-1000}$\,km\,s$^{-1}$. 

\begin{figure}
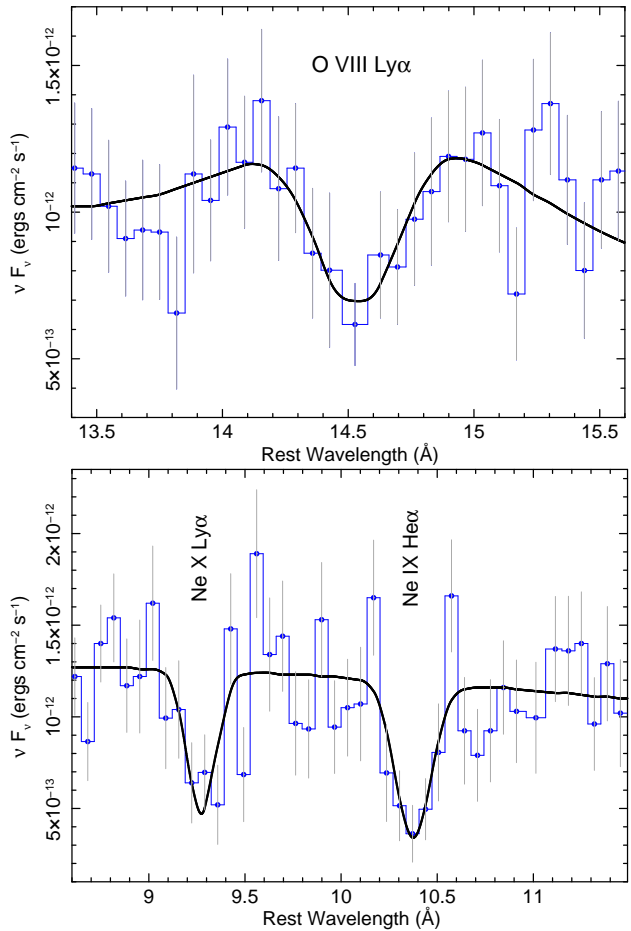

\begin{center}
\rotatebox{-90}{\includegraphics[height=8.7cm]{f3a.eps}}
\rotatebox{-90}{\includegraphics[height=8.7cm]{f3b.eps}}
\end{center}
\caption{Zoom in to the absorption line profiles, with the upper panel showing the putative O\,\textsc{viii} Ly$\alpha$ line and the lower panel showing the $1s\rightarrow2p$ lines from 
Ne\,\textsc{ix} and Ne\,\textsc{x}. The lines are all resolved by the RGS (where one bin corresponds to the FWHM resolution) and can be fitted with a common velocity 
width of $\sigma=3030^{+950}_{-740}$\,km\,s$^{-1}$. Note that the y-axis scale has been converted into energy units using the wavelength conversion given in the text and both the model and data are folded through the instrumental response.}
\label{fig:zoom}
\end{figure}

\subsection{Photoionization Modelling} \label{sec:photoionization_modelling}

To provide a more physically motivated modeling of the absorber, the RGS spectrum was fitted with a grid of photoionized absorption models generated with 
the \textsc{xstar} code \citep{Kallman96}. A grid was adopted for PDS\,456, using the 1--1000\,Rydberg spectral energy distribution defined in an earlier 2017 
{\it XMM-Newton} observation (see \citealt{Reeves18b}). Note that this earlier observation was at a similar X-ray and UV flux level to the 2019b observation and the 
1-1000\,Ryd ionizing luminosity was $L_{\rm ion}=3\times10^{46}$\,erg\,s$^{-1}$. 
A turbulence velocity of 3000\,km\,s$^{-1}$ was chosen for the grid in order to well match the widths of the 
absorption line profiles.

Applying the \textsc{xstar} absorption model to the RGS spectrum gave a good fit, with $\chi^{2}_{\nu}=276.4/251$ and a corresponding improvement in fit statistic of 
$\Delta\chi^{2}=58$ for $\Delta\nu=3$ extra degrees of freedom. The column density of the absorber is $N_{\rm H}=2.3^{+0.9}_{-0.6}\times10^{21}$\,cm$^{-2}$, 
with an ionization parameter of $\log\xi=3.4^{+0.1}_{-0.2}$. The above fit was obtained with the elemental abundances fixed at their Solar values, while the 
improvement in the fit was not significant when we allowed the abundances to vary with respect to the other elements (e.g. for Ne, $\Delta\chi^{2}=2.4$ for $\Delta\nu=1$). 
The outflow velocity was subsequently determined from the \textsc{xstar} fit to be $v/c=-0.258\pm0.003$, in agreement with the simple Gaussian line analysis. 

Note that the line of sight covering fraction of the absorber is $f_{\rm cov}=0.92^{+0.08}_{-0.32}$, thus the minimum 
covering fraction is 60\% and the model is consistent with a fully covering absorber. If the covering fraction were lower than this, then the lines would not be of sufficient depth to model the absorption profiles that are observed in the RGS spectrum.

The resulting fit to the spectrum is shown in the upper panel of Figure~4, while the lower panel shows the \textsc{xstar} model 
with various absorption lines marked at their expected lab-frame wavelengths prior to applying the above velocity shift. 
Comparing the two panels, the three strongest $1s\rightarrow2p$ lines predicted by the model (O\,\textsc{viii} Ly$\alpha$, Ne\,\textsc{ix} He$\alpha$ and 
Ne\,\textsc{x} Ly$\alpha$) are all systematically blue-shifted by the same amount in the observed spectrum when compared to their expected lab-frame wavelengths. 

\subsection{Alternative Line Identifications} \label{sec:identifications}

An identification of the lines was also attempted assuming a zero or a low velocity shift away from their respective rest-frame wavelengths. This corresponds to the scenario whereby the lines originate from a warm absorber, of modest outflow velocity (a few hundred to a few thousand km\,s$^{-1}$), 
as is frequently seen in the soft X-ray spectra of Seyfert 1 galaxies \citep{Kaastra00, Kaspi04, Blustin05, McKernan07}. 

However it is difficult to systematically reproduce the absorption lines without requiring the systematic blueshift. 
For instance, in the scenario whereby the absorber has zero or only a small velocity shift, 
the absorption line observed at 10.37\,\AA\ could be associated to the predicted weak $1s\rightarrow3p$ line of Ne\,\textsc{x} Ly$\beta$, which is expected at a lab-frame wavelength of 10.24\,\AA\ (see Figure 4, lower panel). The observed equivalent width of the 10.37\,\AA\ line is $16\pm5$\,eV. However we 
would then expect to observe the accompanying stronger Ne\,\textsc{x} Ly$\alpha$ line near 12.1\,\AA\ at zero or low velocity. At this wavelength only a tight upper limit of 
$<2$\,eV can be placed on the equivalent width of any absorption line. Thus a zero velocity solution whereby the Ne\,\textsc{x} Ly$\alpha$ transition is at least $\times5$ weaker 
than the corresponding Ly$\beta$ line can be physically ruled out. 

\begin{figure*}
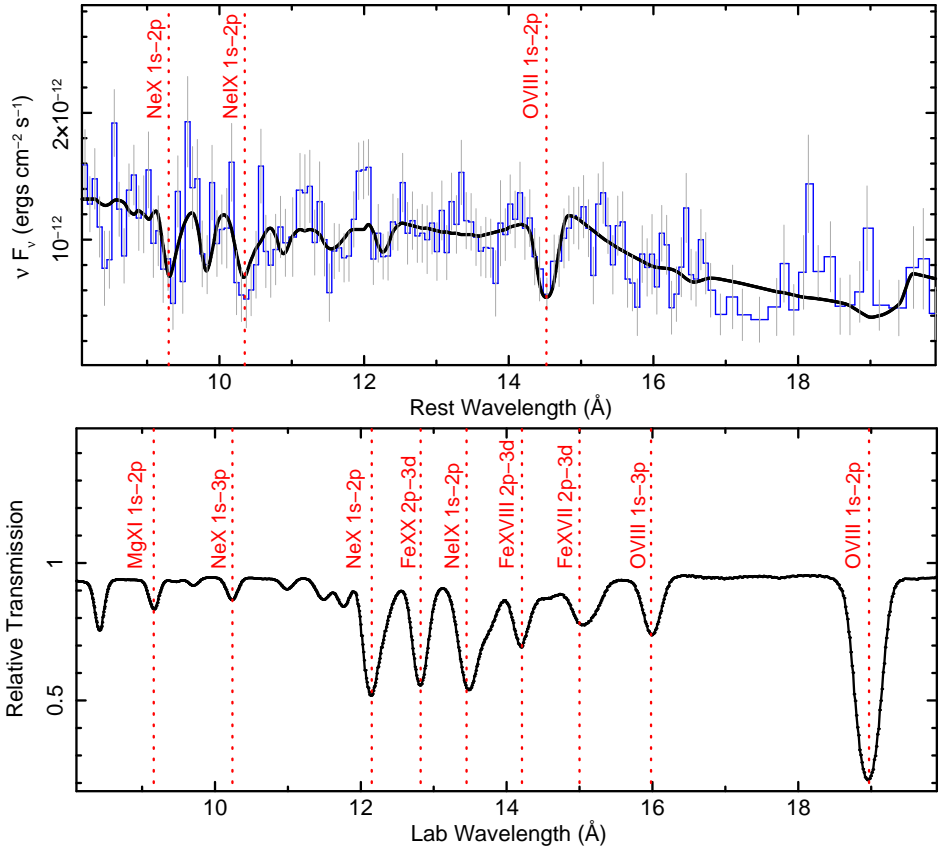

\begin{center}
\rotatebox{-90}{\includegraphics[height=13cm]{f4a.eps}}
\rotatebox{-90}{\includegraphics[height=13cm]{f4b.eps}}
\end{center}
\caption{\textsc{xstar} photoionization model fitted to the 2019b RGS spectrum. The upper panel shows the best-fit model overlayed onto the spectrum, 
which well reproduces the strong $1s\rightarrow2p$ absorption lines from O\,\textsc{viii} Ly$\alpha$, Ne\,\textsc{ix} He$\alpha$ and Ne\,\textsc{x} Ly$\alpha$ 
(as marked by dashed red-lines), with a common velocity shift of $v/c=-0.258\pm0.003$. The lower panel shows the transmission through the same photoionized absorber without 
any velocity shift, i.e. the corresponding lines are at their expected lab-frame wavelengths. The blueshift of the strongest lines in the model relative to their observed 
positions in the upper-panel is apparent. All abundances were fixed at their Solar values. See text for further details.}
\label{fig:xstar}
\end{figure*}

Similarly, without imparting any blue-shift, the observed line at 14.56\,\AA\ falls in between the expected lab wavelengths of the Fe\,\textsc{xvii} and Fe\,\textsc{xviii} L-shell ($2p\rightarrow3d$) transitions (at 15.01\,\AA\ and 14.21\,\AA). Both of these lines are predicted to be weaker (by a 3:1 ratio) in the model spectrum when compared to the strong 
O\,\textsc{viii} Ly$\alpha$ line at 18.97\,\AA, where the latter line is a dominant transition in the photoionized absorber model (Figure~4). 
Instead an upper limit of $EW<3$\,eV can only be placed on any 
18.97\,\AA\ absorption compared to the observed line equivalent of $11.5\pm3.0$\,eV at 14.56\,\AA. Thus, as per the above case of Ne, it appears more plausible for the O\,\textsc{viii} Ly$\alpha$ line to be systematically blue-shifted in order to reproduce the strong absorption profile at 14.56\,\AA.

To quantify this further, the \textsc{xstar} model was refitted to the spectrum, but instead forcing the outflow velocity to be within $\pm10000$\,km\,s$^{-1}$ of zero and allowing its 
column and ionization state to adjust. In this test, the column density drops to a very low value of 
$N_{\rm H}<3.5\times10^{20}$\,cm$^{-2}$ (for an ionization of $\log\xi\sim3$), the model adds little opacity to the spectrum reverting to a simple power-law continuum and the fit statistic is subsequently worse with $\chi^{2}_{\nu}=317.5/251$. Thus the fast wind scenario is instead the likely preferred solution and can consistently account for the 
wavelengths and ratios of the observed lines in the spectrum. Furthermore, as is described below, the velocity of the iron K band absorber is also consistent 
with the soft X-ray gas.

\section{The Broad-Band Spectrum} \label{sec:broadband}

Next the results from the broad-band X-ray spectral analysis of the 2019b observation are presented. 
Data from the EPIC-pn and EPIC-MOS CCD detectors were included over the 0.4--10\,keV observed frame band; 
these have net exposures of 70.1\,ks and 81.0\,ks (after correcting for deadtime) and net count rates of $1.354\pm0.004$\,cts\,s$^{-1}$ and 
$0.697\pm0.003$\,cts\,s$^{-1}$ respectively. The background count rate is about 1\% of the source counts over the whole band and 10\% of the 
source counts over the 7--10\,keV band. 
Spectra from the hard X-ray {\it NuSTAR} FPMA and FPMB detectors were included in the analysis over the 3--40\,keV band, 
with net count rates of $0.044\pm0.001$\,cts\,s$^{-1}$ and $0.042\pm0.001$\,cts\,s$^{-1}$ respectively, while the net exposure per detector was 78.0\,ks. 
Here the background rate is $\sim10$\% of the source rate, although the spectrum becomes background dominated above 40\,keV. 
The {\it NuSTAR} observation, performed in low Earth orbit, overlapped the {\it XMM-Newton} observation, commencing just before and ending just after the {\it XMM-Newton} exposures. A constant multiplicative factor was also included between the {\it NuSTAR} and {\it XMM-Newton} spectra to allow for any cross normalization 
differences between satellites, however this was found to be consistent with 1.0 within errors.

As per the RGS spectral analysis, a neutral Galactic absorption component of $N_{\rm H}=2.4\times10^{21}$\,cm$^{-2}$ was included in the modeling. 
The continuum was modeled by a broken powerlaw, which allows for the fact that the soft X-ray photon index (as seen by RGS) is somewhat steeper than at harder 
X-rays, due to the presence of a soft X-ray excess which is apparent once the Galactic absorber is accounted for. The continuum parameters are reported in Table~3. 
A high energy exponential cut-off was also included, with an e-folding energy fixed to 50\,keV. This is included for consistency with the 2018 
observation which occurred in a bright and flaring state and where the cut-off is well determined at this energy; these results  
will be presented in a later paper. Note the presence of the cut-off does not affect the parameterization of the absorbers which is our focus below. 

\begin{figure}
\begin{center}
\rotatebox{-90}{\includegraphics[height=8.7cm]{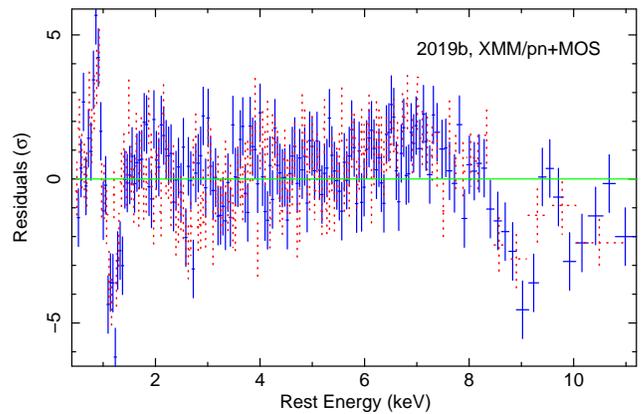}}
\end{center}
\caption{Residuals, plotted in terms of data -- model / error, against a baseline broken power-law continuum to the broad-band {\it XMM-Newton} pn (blue) and MOS (red, dotted points) spectra.  
A strong iron K absorption trough is observed at 9\,keV in the QSO rest frame, which, if it is associated with Fe\,\textsc{xxvi} Ly$\alpha$ at 6.97\,keV, 
requires an outflow velocity of $v/c=-0.25\pm0.02$, consistent with the velocity of the soft X-ray absorber. A second trough may be present at 10.2\,keV, 
which may either be ascribed to a higher velocity wind component or from a contribution from higher order Fe K absorption, as described in the text. 
The presence of the soft X-ray trough near to 1.2\,keV is likely due to an unresolved blend of the Ne K-shell absorption lines, as observed in the RGS.}
\label{fig:del}
\end{figure}

The residuals of the pn and MOS spectra, against this continuum, are plotted in Figure~5 over the 0.5--12\,keV rest frame band. At higher energies, this baseline model 
provides an excellent description of the hard X-ray continuum as seen in {\it NuSTAR}. 
In the iron K band, a strong absorption line is present at rest-frame energy of $\sim 9$\,keV, with a possible second weaker trough near to 10\,keV.
The first line can be modeled with a Gaussian profile of centroid energy $8.98\pm0.13$\,keV, a width of $\sigma=0.31^{+0.14}_{-0.09}$\,keV and an 
equivalent width of $-255^{+58}_{-75}$\,eV. The addition of the line significantly improves the fit statistic by $\Delta\chi^2=-49.6$ for $\Delta\nu=3$ 
change in degrees of freedom. Indeed, the 9\,keV absorption trough has been detected in many of the previous PDS 456 observations 
(e.g. Reeves et al. 2009, Nardini et al. 2015, Matzeu et al. 2017). 
A second line also appears to be significant against the baseline continuum 
($\Delta\chi^2=30.1$ for $\Delta\nu=2$), with $E=10.2\pm0.2$\,keV, an equivalent width of $-246\pm70$\,eV, while the width of the line was assumed to be the same 
as the 9\,keV trough. For completeness, the line parameters are reported in Table\,2.

The absorption features can be identified with the strong resonance lines from Fe\,\textsc{xxv} or Fe\,\textsc{xxvi}, which are frequently detected in the X-ray spectra of many nearby type I AGN \citep{Tombesi10, Gofford13}. 
For the 9\,keV trough, an identification with either of the $1s\rightarrow2p$ lines of Fe\,\textsc{xxv} He$\alpha$ at 6.70\,keV or Fe\,\textsc{xxvi} Ly$\alpha$ at 6.97\,keV returns a velocity shift of $-0.28\pm0.02c$ or $-0.25\pm0.02c$. We note that the latter identification is consistent with the velocity of the soft X-ray outflow. 
In principle, the second 10\,keV trough may be associated with the corresponding higher order ($1s\rightarrow3p$) lines of Fe\,\textsc{xxv} at 7.88\,keV or Fe\,\textsc{xxvi} at 8.25\,keV, 
with a resultant outflow velocity of $-0.25\pm0.02c$ or $-0.21\pm0.02c$ respectively. 
However, the oscillator strengths of these higher order $1s\rightarrow3p$ lines are typically a factor of five weaker than the corresponding strong $1s\rightarrow2p$ 
lines and their equivalent widths are predicted to be weaker by about the same factor. This does not appear to be the case from the observed equivalent widths reported in Table 2. 
Alternatively, the second higher energy absorption trough could originate from a faster wind component, which, if it is associated with the strong Fe\,\textsc{xxvi} 
Ly$\alpha$ line, would then require a second much higher velocity, of $v/c=-0.36\pm0.02$. We investigate this further below in the \textsc{xstar} modeling. 

Interestingly, a third absorption trough is detected in the {\it XMM-Newton} spectra in the soft X-ray band at $E=1.16\pm0.05$\,keV (equivalent to $10.7\pm0.4$\,\AA); this could arise from a blend of the Ne\,\textsc{ix-x} lines seen in the RGS, but which is unresolved at the lower resolution of the CCD spectra.

\subsection{Iron K Absorber Properties} \label{sec:absorber}

To quantify the iron K band absorption, the spectra were again modeled by \textsc{xstar} as above, but adopting a grid of absorption models with a higher turbulence velocity 
of 10000\,km\,s$^{-1}$ to match the widths of the lines measured from their Gaussian profiles. 
Emission from the wind, parameterized by a quasi-spherical distribution of gas, has also been included as per the analysis of \citet{Nardini15}. Note that as detailed by \citet{Nardini15} and \citet{Reeves18b}, the solid angle of the emitter (in terms of $f=\Omega/4\pi$) is determined by the relation, 
$\kappa_{\rm xstar}= f L_{38} / D_{\rm kpc}^2$, where $\kappa_{\rm xstar}$ is the fitted normalization of the \textsc{xstar} emission grid, $L_{38}$ is the 1--1000 Rydberg ionizing 
luminosity in units of $10^{38}$\,ergs\,s$^{-1}$ and $D_{\rm kpc}$ is the distance to PDS\,456 in kpc units (see Table 3 for parameter details). 
The turbulence velocity of the emitter was fixed at 25000\,km\,s$^{-1}$ to account for any broadening of the iron K emission, consistent with previous analysis \citep{Nardini15, 
Reeves18b}. The column density and ionization of the emitter is tied to that of the absorber, i.e. as if the emitting gas is the same matter as the absorber, 
but integrated over a range of angles.  The continuum form is as above.

\begin{figure}
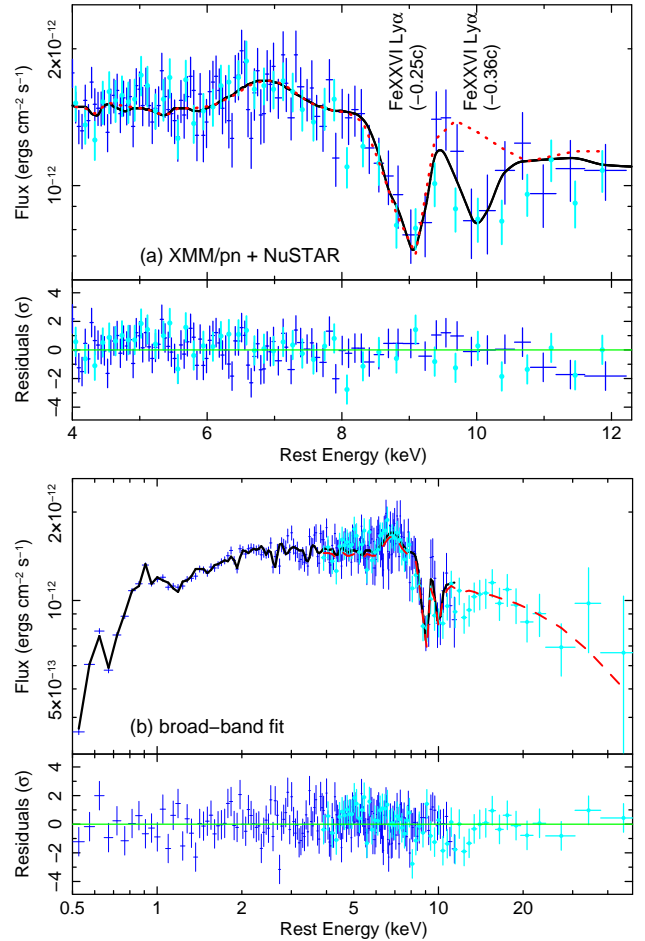

\begin{center}
\rotatebox{-90}{\includegraphics[height=8.7cm]{f6a.eps}}
\rotatebox{-90}{\includegraphics[height=8.7cm]{f6b.eps}}
\end{center}
\caption{Best-fit photoionization model applied to the broad-band {\it XMM-Newton} and {\it NuSTAR} spectra of the 2019b observation. The {\it XMM-Newton} pn spectrum is shown in blue and 
{\it NuSTAR} is in cyan, where for the latter the FPMA and FPMB modules have been combined into a single spectrum. 
The upper panel shows a zoom in around the iron K band, whereby a high ionization photoionized absorber (Table~3, zone 1) can account for the 9\,keV absorption 
trough originating from blueshifted Fe\,\textsc{xxvi} Ly$\alpha$ at 6.97\,keV. Its outflow velocity, of $v/c=-0.261\pm0.007$, is consistent with the soft X-ray absorber seen in the RGS 
and where $v/c=-0.258\pm0.003$. Note that the second absorption trough near to 10\,keV cannot be accounted for by this zone (red dotted line) and requires the addition of a 
second faster wind component with $v/c=-0.360\pm0.013$, which can then fully model both absorption profiles (two velocity model; solid black line).  
The lower panel shows broad-band spectrum and model, whereby here the red dashed line indicates the level of the continuum as fitted to {\it NuSTAR} above 10\,keV.}
\label{fig:broad}
\end{figure}

The resultant fit to the {\it XMM-Newton} and {\it NuSTAR} spectra is shown in Figure~6, where the top panel shows the zoom-in to the iron K band and the lower panel 
the whole broad-band spectrum. The overall fit statistic is $\chi^2=416.2/365$, while the best fit parameters of the model are summarized in Table~3. 
The main component (zone 1) of the Fe K absorber reproduces the 9\,keV absorption trough; its outflow velocity is $v/c=-0.261\pm0.007$, which is in excellent agreement with the RGS absorber where $v/c=-0.258\pm0.003$. Note that its ionization is nearly two orders of magnitude higher than the soft X-ray absorber, 
with $\log\xi=5.0\pm0.1$, while its column density is also correspondingly higher ($N_{\rm H}=7.0^{+1.9}_{-1.6}\times10^{23}$\,cm$^{-2}$). At this degree of 
ionization, most of the opacity arises from H-like iron (e.g. Fe\,\textsc{xxvi} Ly$\alpha$) and the absorber is virtually transparent at soft X-rays, 
with just trace amounts of absorption predicted from H-like Si and S. 

As a result, the broad-band spectrum also requires the addition of the soft X-ray absorber to account for the $\sim 1$\,keV trough. Upon using the same grid as per the RGS analysis, its parameters are found to be consistent in the broad-band spectrum; e.g. $N_{\rm H}=(1.4\pm0.2)\times10^{21}$\,cm$^{-2}$, $\log\xi=3.4^{+0.3}_{-0.1}$ 
and $v/c=-0.25\pm0.02$. This reaffirms the presence of the soft X-ray absorber as measured in the RGS.

Curiously, the zone 1 iron K absorber is not able to account for the second absorption trough near 10\,keV. As noted above, this is mainly because the predicted higher order ($1s\rightarrow3p$) lines of iron are much weaker and make a negligible contribution towards the absorption trough. This is illustrated in the upper panel of Figure\,6, where the red dashed line represents an \textsc{xstar} model with only one velocity component (at $-0.25c$) and which is not able to model the second higher energy trough. 
Instead a second Fe K absorption zone of higher velocity is required, 
with $v/c=-0.360\pm0.013c$ (zone~2, as listed in Table~3), which improves the fit statistic further by $\Delta\chi^2=-43.4$ for $\Delta\nu=2$ and is able to model the 
10\,keV profile (see Figure 6, solid black line for the 2 velocity solution). 
Note that the ionization has been assumed to be the same as per zone\,1.
At this velocity, the 10\,keV trough corresponds to a second velocity component of Fe\,\textsc{xxvi} Ly$\alpha$, perhaps arising from a faster streamline of the wind. 
Note that the presence of more complex velocity 
structure at Fe K has been noted in other observations of PDS\,456 \citep{Reeves09, Reeves18a} and may also be present towards other AGN, e.g. PG 1211+143 \citep{Pounds16}, APM 08279+5255 \citep{Chartas09}, MCG$-$03$-$58$-$007 \citep{Braito18}.  

\begin{deluxetable}{lc}
\tablecaption{Spectral Parameters from Photoionization Modeling to the 2019b observation}
\tablewidth{250pt}
\tablehead{\colhead{Parameter} & \colhead{value}}
\startdata
Fe K absorber, zone 1:-\\
$N_{\rm H}^{a}\times10^{23}$ & $7.0^{+1.9}_{-1.6}$ \\
$\log\xi$$^{b}$ & $5.02\pm0.12$ \\
$v/c$  & $-0.261\pm0.007$ \\
$\Delta\chi^{2}/\Delta\nu$ & $-151.8/2$ \\
\hline
Fe K absorber, zone 2:-\\
$N_{\rm H}^{a}\times10^{23}$ & $4.1^{+3.6}_{-1.5}$ \\
$\log\xi$$^{b}$ & $5.02^{t}$\\
$v/c$  & $-0.360\pm0.013$ \\
$\Delta\chi^{2}/\Delta\nu$ & $-43.4/2$ \\
\hline
Soft X-ray absorber (RGS):-\\
$N_{\rm H}^{a}\times10^{21}$ & $2.3^{+0.9}_{-0.6}$ \\
$\log\xi$$^{b}$ & $3.4^{+0.1}_{-0.2}$ \\
$v/c$ & $-0.258\pm0.003$ \\
$f^c$ & $0.92^{+0.08}_{-0.32}$ \\
$\Delta\chi^{2}/\Delta\nu$ & $-58.0/3$ \\
\hline
Wind Emission:-\\
$N_{\rm H}^{a}\times10^{23}$ & $7.0^t$ \\
$\log\xi$$^{b}$ & $5.02^t$ \\
$\kappa_{\rm xstar}$$^{d}$ & $3.0^{+0.8}_{-0.7}\times10^{-4}$ \\
$\Omega/4\pi$$^{e}$ & $0.53^{+0.14}_{-0.12}$ \\
$\Delta\chi^{2}/\Delta\nu$ & $-45.3/1$ \\
\hline
Continuum:-\\
$\Gamma_{\rm soft}$ & $2.69\pm0.04$ \\
$\Gamma_{\rm hard}$ & $2.03\pm0.03$ \\
$E_{\rm break}$$^{f}$ & $1.24\pm0.05$ \\
$N_{\rm BPL}\times10^{-3}$$^{g}$ & $1.27\pm0.04$\\
$L_{\rm 2-10\,keV}\times10^{44}$$^{h}$ & 2.5 \\
$L_{\rm 1-1000\,Ryd}\times10^{46}$$^{i}$ & 3.0
\enddata
\tablenotetext{a}{Units of column density cm$^{-2}$.}
\tablenotetext{b}{Ionization parameter (where $\xi=L/nR^{2}$) in units of erg\,cm\,s$^{-1}$. Note the ionization of the zone 2 iron K absorber is tied to zone 1.}
\tablenotetext{c}{Covering fraction of absorber, where $f=1$ for a fully covering absorber.}
\tablenotetext{d}{Measured normalization of \textsc{xstar} emission component, where $\kappa_{\rm xstar}= f_{\rm cov} L_{38}/D_{\rm kpc}^2$ and $f_{\rm cov}$ is the emitter covering fraction ($f_{\rm cov}=\Omega/4\pi$), $L_{38}$ is the 1--1000\,Ryd ionizing luminosity in units of $10^{38}$\,erg\,s$^{-1}$ and $D_{\rm kpc}$ is the distance to PDS\,456 in units of kpc.}
\tablenotetext{e}{Solid angle of the emission component ($\Omega/4\pi$), derived from the measured normalization of the emitter as above.}
\tablenotetext{f}{Break energy of broken powerlaw continuum in units of keV. Below the break energy the continuum has a photon index of $\Gamma_{\rm soft}$ 
and above the break energy the photon index is $\Gamma_{\rm hard}$. }
\tablenotetext{g}{Normalization of the broken powerlaw continuum component in units of photons\,cm$^{-2}$\,s$^{-1}$\,keV$^{-1}$ at 1\,keV.}
\tablenotetext{h}{Absorption corrected rest-frame 2--10\,keV luminosity, in units of erg\,s$^{-1}$.}
\tablenotetext{i}{Absorption corrected ionizing luminosity, calculated over the 1-1000\,Rydberg range, in units of erg\,s$^{-1}$.}
\tablenotetext{t}{Denotes parameter is tied.}
\label{tab:xstar}
\end{deluxetable}

\section{Discussion} \label{sec:discussion}

A new observation of PDS\,456 from September 2019 was presented which revealed three resolved lines in the soft X-ray spectrum taken with the {\it XMM-Newton} RGS. 
The lines are required to originate from a fast outflowing absorber, which when fitted with \textsc{xstar} was found to have an outflow velocity of $v/c=-0.258\pm0.003$, with the lines originating from the $1s-2p$ transitions of O\,\textsc{viii} Ly$\alpha$, Ne\,\textsc{ix} He$\alpha$ and Ne\,\textsc{x} Ly$\alpha$. The outflow velocity of the soft X-ray absorber is entirely consistent with the main zone of iron K absorption, confirmed from the same dataset, which has an outflow velocity of $v/c=-0.261\pm0.007$. 

This observation of PDS\,456 represents one of the small, but growing number of cases where the velocity of the ultra fast outflow, as first detected in the iron K band, has been confirmed by an analysis of a grating resolution spectrum in the soft X-ray band (e.g. \citealt{Pounds03,Parker17,Pinto18,Kosec18}).
Indeed, one of the first known objects for the presence of a fast wind, the nearby ($z=0.0809$) QSO PG\,1211+143 \citep{Pounds03}, also simultaneously shows the presence of both a soft X-ray and iron K wind, with a self consistent velocity. The presence of the 
soft X-ray wind in PG\,1211+143, with a velocity of $v/c=-0.06$, was recently confirmed through deep observations both with the {\it XMM-Newton} RGS and {\it Chandra} HETG 
gratings \citep{RLP18, Danehkar18}. This velocity was also consistent with the major component of the iron K absorption \citep{Pounds16}, as well as with the broad Ly$\alpha$ absorption trough as observed from HST/COS spectroscopy \citep{Kriss18}.   

Another notable example, illustrating the co-existence of both soft X-ray and Fe K components of a fast wind, occurs in the NLS1, IRAS 13224$-$3809 (at $z=0.0658$), where the ultra fast outflow also has a typical velocity of $-0.24c$ \citep{Parker17, Pinto18}. Interestingly, from a refined analysis of this AGN, \citet{Chartas18} found a positive correlation between 
the outflow velocity and X-ray luminosity, similar to the behaviour previously seen in PDS\,456 \citep{Matzeu17a} and APM 08279+5255 \citep{Saez11} and which advocated for the affect of radiation pressure on the wind. 

\subsection{Comparison with Previous PDS 456 Observations}

\subsubsection{Comparison with 2018}

The soft X-ray outflows in both PG\,1211+143 \citep{RLP18} and IRAS\,13224$-$3809 \citep{Pinto18} both show pronounced variability, with the outflow opacity generally being higher when the source flux is lower. In the above analyses, this was attributed to variations in either ionization versus flux, or intrinsic variability of the absorber column, or a combination of both. For PDS\,456, the bright 2018 observation performed one year before the 2019b observation 
provides a test case to see how the soft X-ray absorber varies with flux.  
The observed soft X-ray flux in 2018 was $F_{0.4-2.0\,{\rm keV}}=4.8\times10^{-12}$\,ergs\,cm$^{-2}$\,s$^{-1}$ compared to $F_{0.4-2.0\,{\rm keV}}=1.5\times10^{-12}$\,ergs\,cm$^{-2}$\,s$^{-1}$ in the fainter 2019b observation. Note that in the 2-10\,keV band there is a factor of $\times 4$ difference in flux between observations (see Table~1). 
Figure~7 shows the 2018 RGS spectrum with the 2019b spectrum overlayed on the same flux scale.
In contrast to the 2019b spectrum, the 2018 observation is dominated by a bright power-law component, with a harder photon index of $\Gamma=1.90\pm0.04$ and no soft excess 
compared to the 2--10\,keV band, while there are no absorption features apparent in the spectrum. 

\begin{figure}
\begin{center}
\rotatebox{-90}{\includegraphics[height=8.7cm]{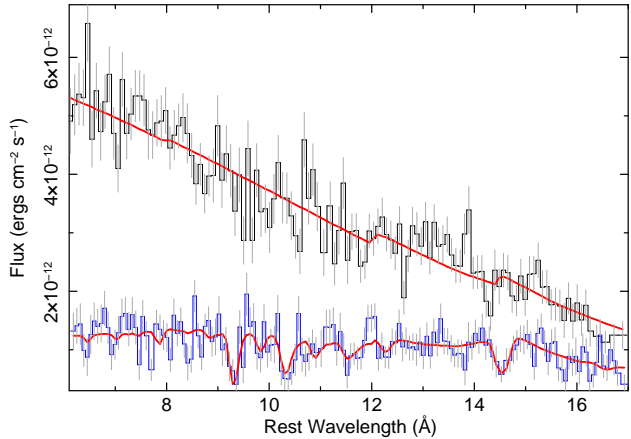}}
\end{center}
\caption{Comparison between the bright 2018 RGS spectrum of PDS\,456 (black points) and the lower flux 2019b spectrum (blue points), with best-fit \textsc{xstar} model over-layed for the latter (red line). With respect to the 2019b observation, 
the flux of the 2018 spectrum was a factor of $\times4$ brighter in both the soft band and the harder 2--10\,keV band. The 2018 spectrum appears featureless, fitted 
by a power-law of $\Gamma=1.90\pm0.04$ modified by Galactic absorption and no significant 
absorption features are apparent close to the wavelengths of the absorption lines detected in the 2019 spectrum. The difference could be reconciled by the subsequent increase in flux which occurred during the 2018 observation, leading to at least a factor of $\times 4$ increase in gas ionization. This is sufficient to render the 2018 
spectrum featureless.}
\label{fig:rgs2018}
\end{figure}

To reconcile this difference, the 2018 spectrum was fitted with the same soft X-ray absorption model as in Section~3.2, allowing 
only the ionization parameter to adjust between the 2018 and 2019b epochs. 
Subsequently, for the same column density, the wind ionization decreased by at least a factor of $\times4$ between the 2018 and 2019b spectra in line with the continuum change; i.e. $\log\xi>3.8$ (in 2018) to $\log\xi=3.3^{+0.1}_{-0.2}$ (in 2019b). As per the above AGN, the featureless nature of the bright 2018 observation is consistent with the wind ionization responding to the continuum flux. Note the change in opacity could equally be expressed in terms of the equivalent factor of $\times4$ change in column density (for 
a constant ionization), where $N_{\rm H}<3.5\times10^{20}$\,cm$^{-2}$ in 2018. 

\subsubsection{Comparison with 2019a}

The 2019b spectrum can also be compared to the 2019a observation which occurred 22 days prior in September 2019 (see Table~1).  As can be seen from Figure\,1, 
its soft X-ray flux was substantially lower (with $F_{0.4-2.0\,{\rm keV}}=9.0\times10^{-13}$\,erg\,cm\,s$^{-1}$). In contrast to the 2018 observation,  
it represents one of the lowest flux observations of PDS\,456; it is comparable to the low flux state observed with {\it Suzaku} in 2013 \citep{Matzeu16}.  

The 2019a spectrum is also shown in Figure~8, where it has been replotted as a ratio the the best-fit continuum model defined from the 2019b observation. 
As per the 2019b observation, a strong absorption trough is present in the Fe K-shell band near 9\,keV and is characterized by a fast, high ionization wind with very similar parameters;  
where $N_{\rm H}=5.0^{+1.4}_{-1.1}\times10^{23}$\,cm$^{-2}$, $\log\xi=5.0\pm0.2$ and $v/c=-0.245\pm0.010$.
In contrast, the 2019a spectrum much is more strongly absorbed in the soft X-ray band, requiring a higher column density of 
$N_{\rm H}=1.24^{+0.08}_{-0.11}\times10^{23}$\,cm$^{-2}$, while its ionization is low with $\log\xi=2.0\pm0.1$. 
The covering fraction of the absorber, against a best-fit power-law continuum of photon index $\Gamma=2.37\pm0.03$, is found to be $f_{\rm cov}=0.77\pm0.02$, 
while the velocity of the soft X-ray absorber is measured to be $v/c=-0.257\pm0.015$. 
The absorption properties of the 2019a observation are summarized in Table~4. The fit statistic with this model is good, with $\chi_{\nu}^2=297.8/302$. 

\begin{figure}
\begin{center}
\rotatebox{-90}{\includegraphics[height=8.7cm]{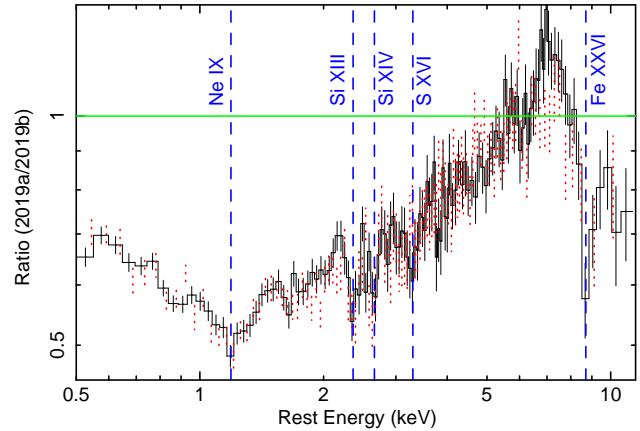}}
\end{center}
\caption{The 2019a {\it XMM-Newton} spectrum of PDS\,456, plotted as a ratio to the continuum model defined in the 2019b observation in Table\,3, 
where the pn is shown in black and MOS in red (dotted lines). 
Compared to the 2019b observation, the 2019a spectrum is much more strongly absorbed in the soft X-ray band and requires a high column density ($N_{\rm H}\sim10^{23}$\,cm$^{-2}$) 
low ionization ($\log\xi=2$) absorber to account for the strong bound-free absorption opacity present. 
As per the 2019b observation (Figure~5), a deep iron K absorption line is present near 9\,keV, with an outflow velocity of $v/c=-0.245\pm0.010$. Note that dashed vertical lines 
represent the possible absorption line structure present in the spectrum, from He/H-like Ne, Si, S and Fe.}
\label{fig:2019a}
\end{figure}

Although the 2019a observation is dominated by the strong bound-free continuum opacity produced by the high column soft X-ray absorber, 
there are some indications of absorption line structure in the soft X-ray spectrum plotted in Figure 8, which are marked by the dashed blue vertical lines. 
In particular, a pronounced trough at 1.2\,keV is observed at the same energy as the 2019b spectrum.  This could also coincide with the Ne\,\textsc{ix-x} absorption blend shifted by $-0.25c$, or 
with some addition contribution from Fe L-shell absorption. 
There are also indications of absorption structures between 2--3.5\,keV, which may be consistent with blueshifted absorption from He and H-like Si and S. 

To investigate this further, we compared the above partially ionized soft X-ray absorber with an alternative neutral partial covering absorber (using the \textsc{zpcfabs} model within \textsc{xspec}), where the outflow velocity is assumed to be zero in the latter model. 
In both cases, we retain the highly ionized iron K wind component in absorption and emission, which reproduces the prominent P Cygni like feature seen in at high energies. 
While the neutral absorber has a similar column density to the ionized one ($N_{\rm H}=1.27^{+0.17}_{-0.14}\times10^{23}$\,cm$^{-2}$ for a covering fraction of $f_{\rm cov}=0.48\pm0.05$), it fails to reproduce the overall shape of the spectrum. 
Figure 9 (left panel) shows the comparison between the ionized and neutral absorbers. 
Significant residuals are present against the neutral absorber, in particular a broad emission and absorption trough near 1 keV and the fit statistic is very poor, with $\chi_{\nu}^2=494.0/304$. In contrast, the ionized absorber is able to account for these soft X-ray residuals, chiefly from the interplay between ionized 
absorption and emission produced from the outflowing gas. 

\begin{figure*}
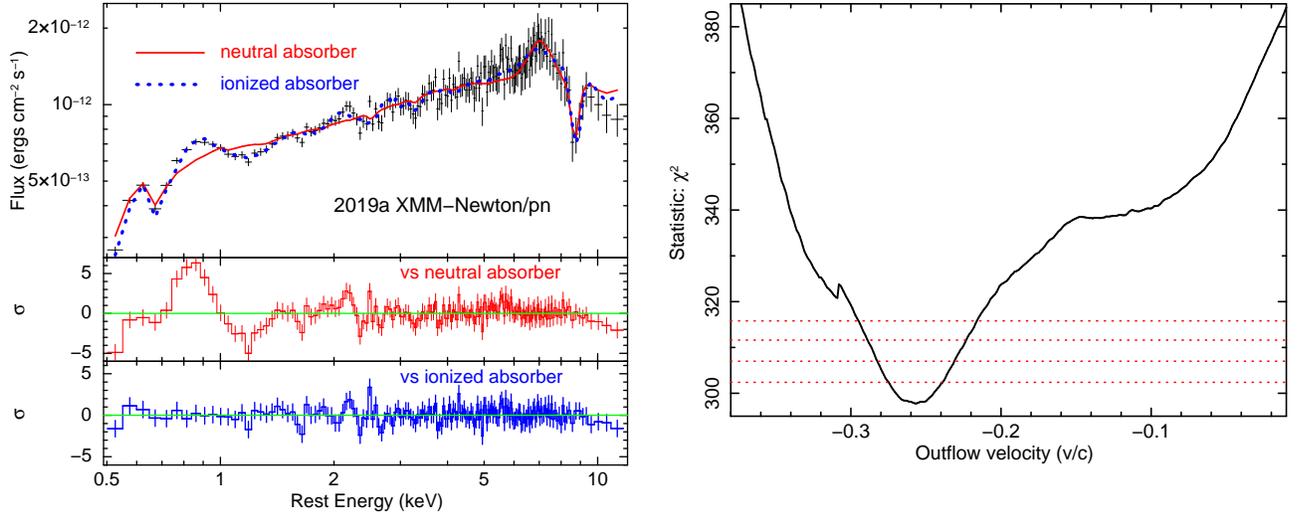

\begin{center}
\rotatebox{-90}{\includegraphics[height=8.7cm]{f9a.eps}}
\rotatebox{-90}{\includegraphics[height=8.7cm]{f9b.eps}}
\end{center}
\caption{Left panel. The 2019a low flux spectrum, compared to either a neutral or partially ionized soft X-ray absorber. The top panel shows the models overlaid on the data, 
where the solid red line corresponds to the neutral absorber and the dotted blue line the ionized absorber. The P Cygni like profile in the iron K band is included in both cases. 
The lower panels show the residuals against these two models. 
Significant soft X-ray residuals are seen against the neutral absorber, which the ionized absorber can account for via a soft X-ray wind. 
Right panel. The fit statistic as a function of outflow velocity for the ionized absorber in the 2019a spectrum, where the dotted horizontal lines correspond to the 90\%, 99\% and 99.9\% and 99.99\% confidence levels for two interesting parameters. The best fit outflow velocity is $v/c=-0.257\pm0.015$, while a lower velocity solution is formally excluded.}
\label{fig:comparison}
\end{figure*}

Note the outflow velocity derived from the ionized soft X-ray absorber, of $v/c=-0.257\pm0.015$, is fully consistent with both the Fe K band absorber and of the 
soft X-ray absorber seen towards the 2019b spectrum. Statistically speaking, the outflow velocity of the soft X-ray absorber is 
well determined, as is illustrated in the right-hand panel of Figure\,9, whereby a lower velocity solution is ruled out with a high degree of confidence (e.g. $\Delta\chi^2>80$ 
for zero outflow velocity). 
Nonetheless the exact velocity determination should be treated with some caution, as the strong spectral curvature towards the 2019a spectrum and its lower resolution precludes a detailed line-by-line analysis, unlike for the 2019b spectrum. In the future, high resolution calorimeter based spectroscopy with \textsc{xrism} and \textsc{athena} 
will be able to provide detailed measurements of these obscuration events.

\begin{deluxetable}{lc}
\tablecaption{Absorption Parameters from the 2019a observation}
\tablewidth{250pt}
\tablehead{\colhead{Parameter} & \colhead{value}}
\startdata
Fe K absorber:-\\
$N_{\rm H}^{a}\times10^{23}$ & $5.0^{+1.4}_{-1.1}$ \\
$\log\xi$$^{b}$ & $5.0\pm0.2$ \\
$v/c$  & $-0.245\pm0.010$ \\
$\Delta\chi^{2}/\Delta\nu$ & $-63.0/3$ \\
\hline
Soft X-ray absorber:-\\
$N_{\rm H}^{a}\times10^{23}$ & $1.24^{+0.08}_{-0.11}$ \\
$\log\xi$$^{b}$ & $2.0\pm0.1$ \\
$v/c$ & $-0.257\pm0.015$ \\
$f^c$ & $0.77\pm0.02$ \\
$\Delta\chi^{2}/\Delta\nu$ & $-288.9/4$ 
\enddata
\tablenotetext{a}{Units of column density cm$^{-2}$.}
\tablenotetext{b}{Ionization parameter (where $\xi=L/nR^{2}$) in units of erg\,cm\,s$^{-1}$.}
\tablenotetext{c}{Covering fraction of absorber, where $f=1$ for a fully covering absorber.}
\label{tab:2019a}
\end{deluxetable}

Overall, the low ionization, high column absorber in the 2019a observation 
produces substantial bound-free opacity and as a result suppresses the continuum towards low energies, with only a fraction 
(about 20\%) of the direct continuum emerging in the soft band. Its effect on the X-ray continuum is similar to what was recently observed in other Seyfert 1s, such as NGC 5548 \citep{Kaastra14}, NGC 3783 \citep{Mehdipour17} and Mrk 335 \citep{Longinotti19, Parker19}, 
where these AGN have undergone an X-ray obscuration event, substantially suppressing the soft X-ray flux. In all these examples, broad and blue-shifted UV absorption line profiles also accompanied the increased X-ray absorption (e.g. \citealt{Kriss19}), linking the obscuration to the emergence of an outflow. The obscuration duration can be pro-longed, lasting years in the cases of NGC\,5548 and Mrk\,335. In PDS\,456, this timescale is likely much shorter, lasting no longer than the 20 day interval between the two 2019 {\it XMM-Newton} 
observations. Indeed similar, short-lived X-ray obscuration events, of the order of days, have also been observed in the Seyfert 1, NGC\,3227 \citep{Turner18} and the QSO, PG\,1211+143 \citep{RLP18}. In the latter AGN, the obscuration was also attributed to an order of magnitude increase in the column of the lower ionization, but fast, soft X-ray wind.

\subsubsection{Comparison with Earlier Observations}

There have also been multiple observations of PDS\,456 with {\it XMM-Newton} prior to the current campaign. These were performed in 2001 \citep{Reeves03}, 2007 \citep{Behar10}, 2013--2014 (5 observations; \citealt{Nardini15, Reeves16}) and 2017 \citep{Reeves18a,Reeves18b}. While a systematic analysis of all of the RGS spectra is beyond the scope of the current paper, a general comparison can be made with these published observations compared to the present epoch. In particular, \citet{Reeves16} 
analyzed all of the archival RGS spectra from 2001--2014. Several of these spectra showed a pronounced broad absorption trough around 1.2\,keV, or in the 
typical wavelength range from 10--12\,\AA. This was attributed to a blend of absorption lines arising from Ne\,\textsc{ix-x} and from L-shell iron, similar to what is also 
predicted here, with an outflow velocity in the range between $0.17-0.27c$, column densities of up to $10^{22}$\,cm$^{-2}$ and ionization parameters of about $\log\xi\sim3.5-4$. Indeed the soft X-ray absorber parameters measured in the 2019b observation are similar to these values, but the column towards PDS\,456 during the low flux 2019a observation is much higher.  

One major difference with 2019b when compared to some of these earlier observations is that the 1.2\,keV absorption trough appeared much broader previously. For example in the 3rd and 4th observations of the 2013--2014 campaign (OBS\,CD, \citealt{Reeves16}), its 
velocity width was $\sigma=28000^{+13000}_{-9000}$\,km\,s$^{-1}$. As a result, it was not possible to resolve the earlier absorption profiles into a distinct series of absorption lines. This has now only been achieved in the current 2019b epoch as the line widths are much narrower, of $\sigma=3000$\,km\,s$^{-1}$, which may indicate a decrease in the turbulence velocity arising from the passage of a different streamline across the line of sight. 
As a result, it was possible in the new 2019b data to both resolve and identify the discrete soft X-ray absorption lines and to confirm that the velocity of both the soft X-ray and iron K absorbers are identical.

\subsection{The Origin of the Soft X-ray Wind}

The major difference between the soft X-ray and iron K wind components in PDS\,456 is that the former has a much lower ionization, by about two orders of magnitude compared to the latter. Thus can the soft X-ray absorber exist co-spatially with the inner high ionization Fe K-shell wind or does it have a separate origin?  
We have also observed the soft X-ray wind column to decrease by almost two orders of in the 20 days between the 2019a and 2019b observations ($N_{\rm H}=10^{23}$\,cm$^{-2}$ to $2\times10^{21}$\,cm$^{-2}$), while its 
ionization also increased by a factor of 10 ($\log\xi\sim2$ to $\log\xi\sim3$).  This is in contrast to the high ionization wind, which remains stable in both column and ionization during this period. Thus at first sight this suggests that the soft X-ray absorber is clumpy, while the high ionization gas is relatively homogeneous.

We first consider the soft X-ray absorber as measured in the 2019b observation. Given that this absorber has a minimum covering fraction towards the X-ray source 
of 60\% (and is consistent with a fully covering absorber), its sizescale must be at least similar to the X-ray source size (if not larger). For PDS\,456, the X-ray coronal size has been previously estimated to be of the order of $\sim10^{15}$\,cm, (or $\sim10 R_{\rm g}$ for a black hole mass of $10^{9}$\,M$_{\odot}$), as measured from X-ray flares 
\citep{Reeves02, Matzeu16, Matzeu17b, Reeves18b}. Thus to achieve a high line of sight coverage, 
the {\it minimum} absorber sizescale is $\Delta R \sim 10^{15}$\,cm, while for a column of $N_{\rm H}\approx10^{21}$\,cm$^{-2}$, 
then the maximum absorber density is $n=N_{\rm H}/\Delta R \sim 10^6$\,cm$^{-3}$. Thus, for an ionizing luminosity of $L_{\rm ion}=3\times10^{46}$\,erg\,s$^{-1}$, an 
ionization parameter of $\log\xi=3.4$ and as $R^{2} = L_{\rm ion} / \xi n$, this yields a minimum radial distance estimate of about a parsec. 
This is also consistent with what was estimated in \citet{Reeves16} for the soft X-ray absorber based on the previous RGS observations of PDS\,456. 

Alternatively, if the soft X-ray absorber were located closer in, at a similar distance to the high ionization wind, then it would be far too compact. At a distance of $R\sim10^{16}$\,cm (or $10^{2} R_{\rm g}$), coincident with the expected launching point of inner iron K disk wind inferred by \citet{Nardini15}, then the soft X-ray absorber density 
is required to be high ($n\sim10^{11}$\,cm$^{-3}$) in order to maintain its lower ionization. 
As a result, its sizescale would then be implausibly small ($\Delta R = N_{\rm H} / n \sim 10^{10}$\,cm). 
In contrast, placing the high ionization absorber at this radius (with $\log\xi=5$, $N_{\rm H}=10^{24}$\,cm$^{-2}$) yields a much more plausible density and sizescale, where 
$n\sim10^{9}$\,cm$^{-3}$ and $\Delta R \sim10^{15}$\,cm. Thus it appears unlikely that the soft X-ray wind can be co-spatial with the innermost disk wind. 

In the 2019a observation, the column of the soft X-ray absorber was much higher at $N_{\rm H}=10^{23}$\,cm$^{-2}$ and its ionization was lower at $\log\xi=2$. 
As above, in order to achieve a large $\sim80$\% line of sight coverage (Table~4), its sizescale must be at least similar to the X-ray source, where $\Delta R \sim 10^{15}$\,cm. 
In this instance, the implied density is $n\sim10^{8}$\,cm$^{-3}$ and thus the derived distance is $R\sim10^{18}$\,cm, i.e. again of the order of a parsec. 
The timescale of the absorber variability also gives an indication of its sizescale.
The absorber variations from 2019a to 2019b occur within a timescale of 1--20 days, i.e. the minimum duration is at least as long as the 2019a {\it XMM-Newton} 
observation, while the absorption has declined by the onset of the 2019b observation. 
If the absorption changes are due to gas moving across our line of sight, then for a maximum transverse cloud velocity of $v=0.25c$, the absorber sizescale is 
constrained to within $\Delta R = v\Delta t \sim 10^{15} - 10^{16}$\,cm, consistent with above. Note the {\it Swift} lightcurve may indicate that the duration of the obscuration event is confined to a few days and thus $\Delta R$ is at the lower end of the range, if it is coincident with the period of low flux at the start of the X-ray monitoring.

The soft X-ray absorption seen towards both the 2019a and b observations is consistent with being at a similar location, they have the same velocity and 
may form part of the same streamline which has a density (and ionization) gradient across the flow. A possible geometry for the absorbing cloud, which may form part of an ensemble, is shown in Figure~9. 
Thus during the heavily obscured 2019a epoch, we are intercepting a denser, higher column portion of the absorber, while 20 days later we are instead viewing through a lower density and more ionized tail of the absorber. This may be similar to what was envisaged by \citet{Maiolino10}, where elongated cometary shaped clouds may be responsible for the short timescale column density and covering fraction variations seen towards the changing look Seyfert, NGC\,1365. Alternatively, a scenario whereby more compact, higher density clouds can intercept the line of sight in front of the lower density gas may also be possible. 

\begin{figure}
\begin{center}
\rotatebox{0}{\includegraphics[width=8.7cm]{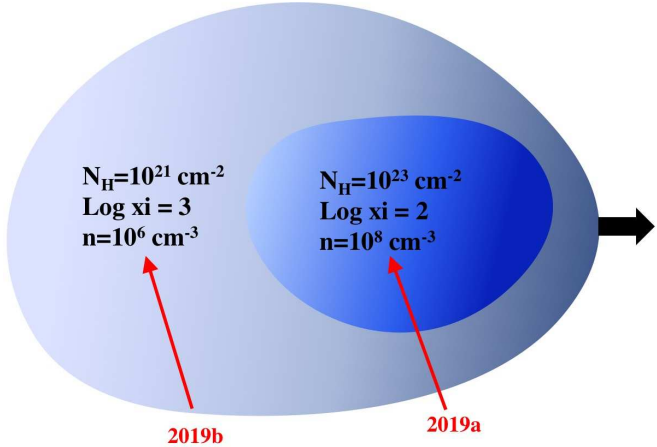}}
\end{center}
\caption{A possible geometry of the soft X-ray absorber, responsible for the variation in column between the 2019a and 2019b observations. 
The red arrows represent our view towards the absorber for the 2019a and 2019b observations and the thick black arrow the direction of the cloud. 
During the 2019a observation, our sightline intercepts a denser, higher column ($10^{23}$\,cm$^{-2}$) part of the absorber (dark blue), while in 2019b the absorption has decreased as we now view through lower density and lower column gas (light blue). Note that the maximum sizescale of the absorber is set by the 20 day timescale between observations ($\Delta R\sim10^{16}$\,cm), while to maintain its lower ionization, the absorber is likely located on parsec scales.}
\label{fig:clump}
\end{figure}

In PDS\,456, the clumpy, lower ionization soft X-ray absorbing clouds likely form from gas further out (on parsec scales) compared to the highest ionization phase of the disk wind, which is launched closer to the black hole. A similar scenario was postulated by \citet{Serafinelli19} for PG\,1114+445, whereby a possible fast soft X-ray 
absorber originates from clouds entrained within the high ionization UFO. 

We can consider further the mass outflow rate of the soft X-ray wind and the subsequent geometric covering fraction ($f_{\rm cov} = \Omega/4\pi$) and volume filling factor ($f_{\rm v}$) of the gas. Taking an expression for the mass outflow rate ($\dot{M}_{\rm out}$) which accounts for the covering and filling factors of the gas (e.g. \citealt{Blustin05}), then:-

\begin{equation}
\dot{M}_{\rm out} \approx 4\pi f_{\rm cov} f_{\rm v} \mu m_{\rm p} v_{\rm out} n R^{2},
\end{equation}

\noindent where $\mu=1.3$ for Solar abundances and $nR^{2} = L_{\rm ion} / \xi$ from the definition of the ionization parameter. For the soft X-ray wind in PDS\,456, where 
$L_{\rm ion} = 3\times10^{46}$\,ergs\,s$^{-1}$ and $\log\xi=3.4$, then the product $nR^2 = 10^{43}$\,cm$^{-1}$, while $v_{\rm out}=0.25c$. 
Thus the soft X-ray mass outflow rate is $\dot{M}_{\rm out} \approx10^4 f_{\rm cov} f_{\rm v}$\,M$_{\odot}$\,yr$^{-1}$. 
This seems implausibly high for a homogeneous outflow, where $f_{\rm v}\sim1$ and would easily exceed the expected mass accretion rate for PDS\,456.

In contrast, for the innermost, high ionization ($\log\xi=5$) iron K wind component, originating from close to the launch radius ($R\sim100R_{\rm g}$ or $\sim10^{16}$\,cm), \citet{Nardini15} 
derived a mass outflow rate of $\dot{M}_{\rm out}\approx 10$\,M$_{\odot}$\,yr$^{-1}$. Such highly ionized gas is likely to be more homogeneous in nature (i.e. $f_{v}\sim1$), 
while \citet{Nardini15} derived a high geometric covering factor for the gas of $f_{\rm cov}\sim0.5$, as measured by its iron K-shell P-Cygni profile. 
Instead, while the soft X-ray mass outflow rate cannot be reliably estimated per se, the above argument supports the hypothesis whereby this gas is highly clumped, where its volume filling factor is 
likely to be at least $f_{\rm v}=10^{-3}$. This is also consistent with the above sizescales and radial distance for the soft X-ray absorber, where one would typically estimate that $\Delta R / R \sim 10^{-3}$. 

Interesting the heavily obscured 2019a {\it XMM-Newton} observation commenced just two days prior to an observation with HST/COS, where the aim of the latter was to measure the properties of any UV outflow towards PDS 456 and its relation to the X-ray wind. Indeed, the high column density ($N_{\rm H}=10^{23}$\,cm$^{-2}$) and low ionization ($\log\xi=2$) of the X-ray absorbing gas should leave its imprint on the UV spectrum, producing broad absorption lines such as from C\,\textsc{iv}, as has also been discussed by \citet{Hamann18} for PDS\,456. 
The depth of any such UV troughs would critically depend on the geometry and covering factor of the absorbing clouds, relative to a more extended UV continuum source compared to the  compact X-ray corona. For instance, if the clouds are relatively compact (i.e. $\Delta R\sim10^{15}$\,cm) and have a low filling factor, then they may only obscure a small portion of the near UV continuum. Future work (Hamann et al. in prep) will quantify the presence of any UV absorption towards PDS\,456 during the 2019a epoch and the implications for properties of the obscuring gas.

In feedback scenarios, such as those suggested by \citet{King10}, the soft X-ray absorber may arise from the slower post-shock gas, resulting from the interaction of the inner ultra fast outflow with the surrounding ambient ISM. This was suggested to be a possible origin of the soft X-ray wind in IRAS\,17020+4544 \citep{Longinotti15}, where in particular the complex velocity behaviour of the slower warm absorbing gas may result from the interaction of the wind with the ambient ISM \citep{Sanfrutos18}. 
Likewise, in NGC\,4051, a wealth of absorption lines with velocity components of up to $\sim -10\,000$\,km\,s$^{-1}$ were detected \citep{PoundsVaughan11}, which were interpreted as originating from a cooling shocked flow \citep{PoundsKing13}. 
However this cannot be the case in PDS\,456, as the soft X-ray clouds have the same velocity as the Fe K-shell wind and are continuing to coast at the terminal velocity of $-0.25c$. 

A low filling factor of the clumpy soft X-ray gas could also impact the efficiency of the X-ray wind for imparting energy conserving feedback on larger scales.
For instance, from the CO observations of PDS\,456 obtained from ALMA, \citet{Bischetti19} showed that the kinetic power of the larger scale molecular outflow is likely much lower than the inner X-ray wind, by a typical factor of $10^{-2} - 10^{-3}$. In principle, the inhomogeneous nature of the X-ray wind may account for the relatively low efficiency factor in both this and towards other AGN \citep{Fiore17, Bischetti19, Sirressi19, RB19}. 

\section{Acknowledgements}

JR acknowledges financial support through grants 80NSSC18K1603 and HST-GO-14477. VB acknowledges support from grant number GO7-18091X. 
VB and JR both acknowledge support from grant 80NSSC20K0793. 
EN acknowledges financial contribution from the agreement ASI-INAF n.2017-14-H.0 and partial support from the EU Horizon 2020 Marie Sk\l{}odowska-Curie grant agreement no. 664931
FH acknowledges support from the Hubble Space Telescope guest observer program HST-GO-15309. 
Based on observations obtained with XMM-Newton, an ESA science mission with instruments and contributions directly funded by ESA Member States and NASA.

\end{document}